\documentclass[12pt]{article}
\title{The Classical Gravitational $N$-Body Problem}
\author{Douglas C. Heggie\\
University of Edinburgh,\\
School of Mathematics,\\
King's Buildings,\\
Edinburgh EH9 3JZ,\\
UK}
\date{}
\def\br{\mathbf r}
\def\bv{\mathbf v}
\begin{document}
\maketitle


\section{Introduction}

Let a number, $N$, of particles interact classically
through Newton's Laws of Motion and Newton's  inverse square Law
of Gravitation.  Then the equations of motion are
\begin{equation}\label{eq:eom}
\ddot\br_i = - G\sum_{j = 1,j\ne i}^{j=N}m_j\frac{\br_i - \br_j}{\vert\br_i - \br_j\vert^3}.
\end{equation}
where $\br_i$ is the position vector of the $i$th particle relative to
some inertial frame, $G$ is the universal constant of gravitation, and
$m_i$ is the mass of the $i$th particle.  These equations provide an
approximate mathematical model with numerous applications in
astrophysics, including the motion of the moon and other bodies in the
Solar System (planets, asteroids, comets and meteor particles); stars
in stellar systems ranging from binary and other multiple stars to
star clusters and galaxies; and the motion of dark matter particles in
cosmology.  For $N=1$ and $N=2$ the equations can be solved
analytically.  The case $N=3$ provides one of the richest of all
unsolved dynamical problems -- the general three-body problem.  For
problems dominated by one massive body, as in many planetary problems,
approximate methods based on perturbation expansions have been
developed.  In stellar dynamics, astrophysicists have developed
numerous numerical and theoretical approaches to the problem for
larger values of $N$, including treatments based on the Boltzmann
equation and the Fokker-Planck equation; such $N$-body systems can
also be modelled as self-gravitating gases, and thermodynamic insights
underpin much of our qualitative understanding.

\section{Few-Body Problems}
\subsection{The two-body problem}
For $N=2$ the relative motion of the two bodies can be reduced to the
force-free motion of the centre of mass and the problem of the
relative motion.  If $\br = \br_1-\br_2$, then
\begin{equation}\label{eq:kepler}
\ddot\br = - G(m_1+m_2)\frac{\br}{\vert\br\vert^3},
\end{equation}
often called the Kepler Problem.  It represents motion of a particle
of unit mass
under a central inverse-square force of attraction.  Energy and
angular momentum are constant, and the motion takes place in a plane
passing through the origin.  Using plane polar coordinates
$(r,\theta)$ in this plane, the equations for the energy and angular
momentum reduce to 
\begin{eqnarray}
E &=& \frac{1}{2}\left({\dot r}^2 + \frac{L^2}{r^2}\right) - \frac{G(m_1+m_2)}{r}\label{eq:E}
\\
\label{eq:L}
L &=& r^2\dot\theta. 
\end{eqnarray}
(Note that these are not the energy and angular momentum of the
two-body problem, even in the barycentric frame of the centre of mass;
$E$ and $L$ must be multiplied by the reduced mass
$m_1m_2/(m_1+m_2)$.)  Using eqs.(\ref{eq:E}) and (\ref{eq:L}) the
problem is reduced to quadratures.  The solution shows that the motion
is on a conic section (ellipse, circle, straight line, parabola or
hyperbola), with the origin at one focus.

This reduction depends on the existence of integrals of the equations
of motion, and these in turn depend on symmetries of the underlying
Lagrangian or Hamiltonian.  Indeed eqs.(\ref{eq:eom}) yield ten first
integrals: six yield the rectilinear motion of the centre of mass,
three the total angular momentum and one the energy.  Furthermore,
eq.(\ref{eq:kepler}) may be transformed, via the Kustaanheimo-Stiefel
transformation, to a four-dimensional simple harmonic oscillator.
This reveals further symmetries, corresponding to further invariants:
the three components of the Lenz vector.  Another manifestation of the
abundance of symmetries of the Kepler problem is the fact that there
exist action-angle variables in which the Hamiltonian depends on only
one action, i.e. $H = H(L)$.   Another application of the  KS
transformation is one that  has practical
importance: it removes the singularity of (i.e. regularises) the Kepler problem at
$r=0$, which is troublesome numerically.

To illustrate the character of the KS transformation, we consider
briefly the planar case, which can be handled with a complex variable
obeying the equation of motion $\ddot z = - z/\vert z\vert^3$ (after
scaling eq.(2)).  By introducing the Levi-Civita transformation $z =
Z^2$ and Sundman's transformation of the time, i.e. $dt/d\tau = \vert
z\vert$, the equation of motion transforms to $Z'' = hZ/2$, where $h =
\vert\dot z\vert^2/2 - 1/\vert z\vert$ is the constant of energy.  The
KS transformation is a very similar exercise using quaternions.

\subsection{The restricted three-body problem}
The simplest three-body problem is given by the motion of a test
particle in the gravitational field of two particles, of positive mass
$m_1, m_2$, in 
circular Keplerian motion.  This is called the circular restricted
three-body problem, and the two massive particles are referred to as
primaries.  In a rotating frame of reference, with origin at the
centre of mass of these two particles,  which are at rest at positions $\br_1, \br_2$, the equation of
motion is
\begin{equation}\label{eq:rtbp}
\ddot\br + 2\Omega\times\dot\br + \Omega\times(\Omega\times\br) =
G\nabla\left(\frac{m_1}{\vert\br-\br_1\vert} + \frac{m_2}{\vert\br-\br_2\vert}\right),
\end{equation}
where $\br$ is the position of the massless particle and $\Omega$ is
the angular velocity of the frame.

This problem has three degrees of freedom but only one known integral:
it is the Hamiltonian in the rotating frame, and equivalent to the Jacobi
integral, $J$.  One consequence is that 
Liouville's theorem is not applicable, and more elaborate arguments
are required to decide its integrability.  Certainly, no general
analytical solution is known.  

\begin{figure}
\vskip4truein
\includegraphics{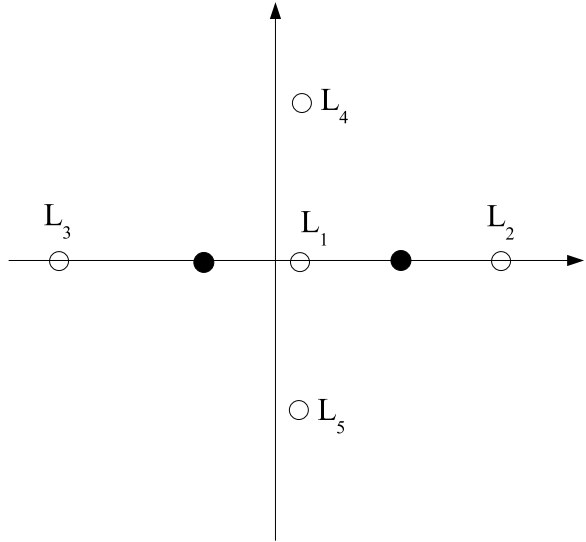}
\caption{The equilibrium solutions of the circular restricted
three-body problem.  We choose a rotating frame of reference
in which two particles are at rest on the $x$-axis.  The massless
particle is at equilibrium at each of the five points shown.  Five
similar configurations exist for the general three-body problem; these
are the ``central'' configurations.}
\label{fig:eulerlagrange}
\end{figure}

There are five {\sl equilibrium} solutions,
discovered by Euler and Lagrange (see Fig.\ref{fig:eulerlagrange}).
They lie at critical points of the effective potential in the rotating
frame, and demarcate possible regions of motion.

Throughout the twentieth century, much numerical effort was used in
finding and classifying {\sl periodic} orbits, and in determining their
stability and bifurcations.  For example there are families of
periodic orbits close to each primary; these are perturbed Kepler
orbits, and are referred to as satellite motions.  Other important families are the series of Liapounov orbits
starting at the equilibrium points.

Some variants of the restricted three-body problem include
\begin{enumerate}
\item the {\sl elliptic} restricted three-body problem, in which the
primaries move on an elliptic Keplerian orbit;  in suitable
coordinates the equation of motion closely resembles
eq.(\ref{eq:rtbp}), except for a factor on the right side which
depends explicitly on the independent variable (transformed time);
this system has no first integral.
\item {\sl Sitnikov's} problem, which is a special case of the elliptic
problem, in which $m_1=m_2$, and the motion of the massless particle
is confined to the axis of symmetry of the Keplerian motion;  this is
still non-integrable, but simple enough to allow extensive analysis of
such fundamental issues as integrability and stochasticity;
\item {\sl Hill's} problem, which is a scaled version suitable for examining
motions close to one primary; its importance in applications began
with studies of the motion of the moon, and it remains vital for
understanding the motion of asteroids.
\end{enumerate}

\subsection{The general three-body problem}

\subsubsection{Exact solutions}
When all three particles have non-zero masses, the equations of motion
become
\[
m_i\ddot \br_i = - \nabla_iW,
\]
where the potential energy is 
\[
W = - G\sum_{1\le i<j\le3}\frac{m_im_j}{\vert\br_{i}-\br_j\vert}.
\]
Then the exact solutions of
Euler and Lagrange survive in the form of {\sl homographic} solutions.
In these solutions the configuration remains geometrically similar,
but may rotate and/or pulsate in the same way as in the two-body
problem.

  Let us represent the position vector $\br_i$ in the planar
three-body problem by the complex number $z_i$.  Then it is easy to
see that we have a solution of the form $z_i(t) = z(t)z_{0i}$,
provided that 
\[
\ddot z = -C\frac{z}{\vert z\vert^3}
\]
and
\[
m_iCz_{0i} = \nabla_iW(z_{01},z_{02},z_{03}),
\]
for some constant $C$.  Thus $z(t)$ may take the form of any solution
of the Kepler problem, while the complex numbers $z_{0i}$ must
correspond to what is called a {\sl central configuration}.  These are
in fact critical points of the scale-free function $W\sqrt{I}$, where
$I$ (the ``moment of inertia of the system'') is given by $I =
\sum_1^3 m_ir_i^2$; and $C = - W/I$.

\begin{figure}
\vskip4truein
\includegraphics{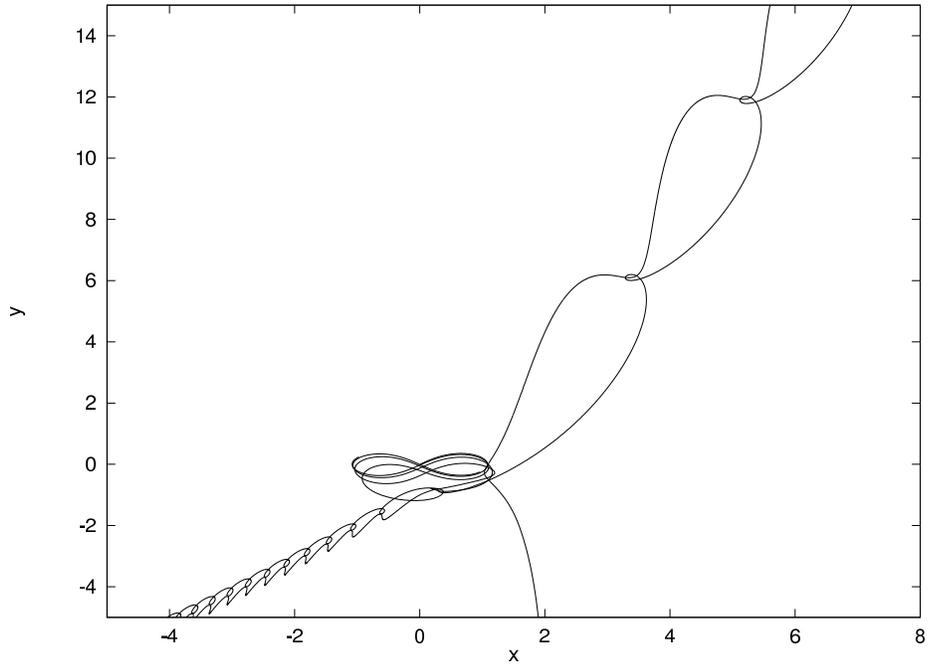}
\caption{A rare example of a scattering encounter between two binaries
(which approach from upper right and lower left) which leads to a
permanently bound triple system describing the ``figure 8'' periodic
orbit.  A fourth body escapes at the bottom.  Note the differing
scales on the two axes. {\sl Originally published in MNRAS}}
\label{fig:eight}
\end{figure}

The existence of other important classes of periodic solutions can be
proved analytically, even though it is not possible to express the
solution in closed form.  Examples include {\sl hierarchical}
three-body systems, in which two masses $m_1, m_2$ exhibit nearly
elliptic relative motion, while a third mass orbits the barycentre of
$m_1$ and $m_2$ in another nearly elliptic orbit.  In the mathematical
literature this is referred to as motion of {\sl elliptic-elliptic}
type.  More surprisingly, the existence of a periodic solution in
which the three bodies travel in succession along the same path,
shaped like a figure 8 (cf. fig\ref{fig:eight}), was established by Chenciner \& Montgomery
(2000), following its independent discovery by Moore using numerical
methods.  Another interesting periodic motion that was discovered
numerically, by Schubart, is a solution of the {\sl collinear}
three-body problem, and so collisions are inevitable.  In this motion
the body in the middle alternately encounters the other two bodies.

\subsubsection{Singularities}

As Schubart's solution illustrates, two-body encounters can occur in
the three-body problem.  Such singularities can be regularised just as
in the pure two-body problem.  Triple collisions can not be
regularised in general, and this singularity has been studied by the
technique of ``blow-up''.  This has been worked out most thoroughly in
the collinear three-body problem, which has only two degrees of
freedom.  The general idea is to transform to two variables, of which
one (denoted by $r$, say) determines the scale of the system, while
the other ($s$) determines the configuration (e.g. the ratio of
separations of the three masses).  By scaling the corresponding
velocities and the time, one obtains a system of three equations of
motion for $s$ and the two velocities which are perfectly regular in
the limit $r\to 0$.  In this limit, the energy integral restricts the
solutions of the system to a manifold (called the {\sl collision
manifold}).  Exactly the same manifold results for zero-energy
solutions, which permits a simple visualisation.  
Equilibria on the collision manifold correspond to the Lagrangian
collinear solutions in which the system either expands to infinity or
contracts to a three-body collision.  

\subsubsection{Qualitative ideas}

Reference has already been made to motion of elliptic-elliptic type.
In motion of {\sl elliptic-hyperbolic} type there is again an
``inner'' pair of bodies describing nearly Keplerian motion, while the
relative motion of the third body is nearly hyperbolic.  In
applications this is referred to as a kind of {\sl scattering}
encounter between a binary and a third body.  When the encounter is
sufficiently close, it is possible for one member of the binary to be
exchanged with the third body.  One of the major historical themes of
the general three-body problem is the classification of connections
between these different types of asymptotic motion.  It is possible to
show, for instance, that the measure of initial conditions of
hyperbolic-elliptic type leading asymptotically to elliptic-elliptic
motion (or any other type of permanently bound motion) is zero.  Much
of the study of such problems has been carried out numerically.

There are many ways in which the {\sl stability} of three-body motions may
be approached.  One example is furnished by the central configurations
already referred to.  They can be used to establish sufficient
conditions for ensuring that exchange is impossible, and similar conclusions.

A powerful tool for qualitative study of three-body motions is Lagrange's
identity, which is now thought of as the reduction to three bodies of
the {\sl virial theorem}.  Let the size of the system be characterised
by the ``moment of inertia'' $I$.  Then  it is easy to show that
$$
\frac{d^2I}{dt^2} = 4T + 2W,
$$
where $T, W$ are, respectively, the kinetic and potential energies of
the system.  Usually the barycentric frame is adopted.  Since $E =
T+V$ is constant and $T\ge0$, it follows that the system is not
bounded for all $t>0$ unless $E<0$.  

\subsubsection{Perturbation theory}

The question of the integrability of the general three body problem has stimulated much
research, including the famous study by Poincar\'e which established
the non-existence of integrals beyond the ten classical ones.
Poincar\'e's work was an important landmark in the application to the
three-body problem of {\sl perturbation methods}.  If one mass
dominates, i.e. $m_1\gg m_2$ and $m_1\gg m_3$, then the motion of
$m_2$ and $m_3$ relative to $m_1$ is mildly perturbed two-body motion,
unless $m_2$ and $m_3$ are close together.  Then it is beneficial to
describe the motion of $m_2$ relative to $m_1$ by the parameters of
Keplerian motion.  These would be constant in the absence of $m_3$,
and vary slowly because of the perturbation by $m_3$.  This was the
idea behind Lagrange's very general method of variation of parameters
for solving systems of differential equations.  Numerous methods were
developed for the iterative solution of the resulting equations.  In
this way the solution of such a three-body problem could be
represented as a type of trigonometric series in which the arguments
are the angle variables describing the two approximate Keplerian
motions.  These were of immense value in solving problems of celestial
mechanics, i.e. the study of the motions of planets, their satellites,
comets and asteroids.

A major step forward was the introduction of Hamiltonian
methods.  A three-body problem of the type we are considering has a
Hamiltonian of the form 
\[
H = H_1(L_1) + H_2(L_2) + R,
\]
where $H_i, i = 1,2$ are the Hamiltonians describing the interaction
between $m_i$ and $m_1$, and $R$ is the ``disturbing function''.  It
depends on all the variables, but is small compared with the $H_i$.
Now perturbation theory reduces to the task of performing canonical
transformations which simplify $R$ as much as possible.  

Poincar\'e's major contribution in this area was to show that the
series solutions produced by perturbation methods are not, in general,
convergent, but asymptotic.  Thus they were of practical rather than
theoretical value.  For example, nothing could be proved about the
stability of the solar system using perturbation methods.  It took the
further analytic development of KAM theory to rescue this aspect of
perturbation theory.  This theory can be used to show that, provided
that two of the three masses are sufficiently small, then for almost
all initial conditions the motions remain close to Keplerian for all
time.  Unfortunately now it is the practical aspect of the theory
which is missing;  though we have introduced this topic in the context
of the three-body problem, it is extensible to any $N$-body system
with $N-1$ small masses in nearly-Keplerian motion about $m_1$, but to
be applicable to the solar system the masses of the planets would have
to be ridiculously small.

\subsubsection{Numerical methods}

Numerical integrations of the three-body problem were first carried
out near the beginning of the 20th century, and are now commonplace.
For typical scattering events, or other short-lived solutions, there
is usually little need to go beyond common Runge-Kutta methods, provided
that automatic step-size control is adopted.  When close two-body
approaches occur, some regularisation based on the KS transformation
is often exploited.  In cases of prolonged elliptic-elliptic motion,
an analytic approximation based on Keplerian motion may be adequate.
Otherwise (as in problems of planetary motion, where the evolution
takes place on an extremely long time scale), methods of very high
order are often used.  Symplectic methods, which have been developed
in the context of Hamiltonian problems, are increasingly adopted for
problems of this kind, as their long-term error behaviour is generally
much superior to that of methods which ignore the geometrical
properties of the equations of motion.

\subsection{Four- and five-body problems}

Many of the foregoing remarks, on central configurations, numerical
methods, KAM theory, etc, apply equally to few-body problems with
$N>3$.  Of special interest from a theoretical point of view is the
occurrence of a new kind of singularity, in which motions become
unbounded in finite time.  For $N=4$ the known examples also require
two-body collisions, but non-collision orbits exhibiting finite-time
blow-up are known for $N=5$. 

 One of the practical (or, at least,
astronomical) applications is again to scattering encounters, this
time involving the approach of two binaries on a hyperbolic relative
orbit.  Numerical results show that a wide variety of outcomes is
possible, including even the creation of the figure-8 periodic orbit
of the three-body problem, while a fourth body escapes (Fig.\ref{fig:eight}).

\section{Many-Body Problems}

Many of the concepts already introduced, such as the virial theorem,
apply equally well to the many-body classical gravitational problem.
In this section we refer mainly to the new features which arise when
$N$ is not small.  In particular, statistical descriptions become
central.  The applications also have a different emphasis, moving from
problems of planetary dynamics (celestial mechanics) to those of
stellar dynamics.  Typically, $N$ lies in the range $10^2$--$10^{12}$.

\subsection{Evolution of the distribution function}

The most useful statistical description is obtained if we neglect
correlations and focus on the one-particle distribution function
$f(\br,\bv,t)$, which can be interpreted as the number-density at time
$t$ at the
point in phase space corresponding to position $\br$ and velocity
$\bv$.  Several processes contribute to the evolution of $f$.

\subsubsection{Collective effects}

When the effects of near neighbours are neglected,  the dynamics is
described by the {\sl Vlasov-Poisson} system
\begin{eqnarray}
\label{eq:vlasov}
\frac{\partial f}{\partial t} + \bv.\frac{\partial f}{\partial\br} -
\frac{\partial\phi(\br,t)}{\partial\br}.\frac{\partial f}{\partial\bv} &=& 0
\\
\label{eq:Poisson}
\nabla^2\phi &=& 4\pi Gm\int f(\br,\bv,t)d^3\bv,
\end{eqnarray}
in which $\phi$ is the gravitational potential, and $m$ is the mass of
each body.  Obvious extensions are necessary if
not all bodies have the same mass.  

Solutions of eq.(\ref{eq:vlasov}) may be found by the
method of characteristics, which is most useful in cases where the
equation of motion $\ddot\br = -\nabla\phi$ is integrable, e.g. in
stationary, spherical potentials.  An example is the solution 
\begin{equation}
\label{eq:plummer}
f =
\vert E\vert^{7/2},
\end{equation}
 where $E$ is the specific energy of a body, i.e. $E
= v^2/2 + \phi$.  This satisfies eq(\ref{eq:vlasov}) provided that
$\phi$ is static.  Eq.(\ref{eq:Poisson}) is satisfied provided that
$\phi$ satisfies a case of the Lane-Emden equation, which is easy to
solve in this case.  

The solution just referred to is an example of an {\sl equilibrium} solution.
In an equilibrium solution the
virial theorem takes the form $4T + 2W = 0$, where $T, W$ are
appropriate mean-field approximations for the kinetic and potential
energy, respectively.  It follows that $E = -T$, where $E = T + V$ is
the total energy.  An increase in $E$ causes a decrease in $T$, which
implies that a self-gravitating $N$-body system exhibits a negative
specific heat.

There is little to choose between one
equilibrium solution and another, except for their stability.  In such
an equilibrium, the bodies orbit within the potential on a timescale
of the {\sl crossing time}, which is conventionally defined to be
$t_{cr} = \displaystyle{\frac{GM^{5/2}}{(2\vert E\vert)^{3/2}}}$.

The most important evolutionary phenomenon of collisionless dynamics is {\sl
violent relaxation}.  If $f$ is not time-independent then $\phi$ is
time-dependent in general.  Also, from the equation of motion of one
body, $E$ varies according to $dE/dt = \partial\phi/\partial
t$, and so energy is exchanged between bodies, which leads to an evolution of
the distribution of energies.  This process is known as violent
relaxation.  

Two other relaxation processes are of importance:
\begin{enumerate}
\item Relaxation is  possible on each energy hypersurface,
even in a static potential, if the potential is non-integrable. 
\item  The range of collective phenomena becomes remarkably rich if the
system exhibits ordered motions, as in rotating systems.  Then an
important role is played by resonant motions, especially resonances of low
order.  The corresponding theory lies at the basis of the theory of
spiral structure in galaxies, for instance.
\end{enumerate}

\subsubsection{Collisional effects}

The approximations of collisionless stellar dynamics suppress two
important processes:
\begin{enumerate}
\item the exponential divergence of
stellar orbits, which 
takes place on a time scale of order $t_{cr}$.  Even in an integrable
potential, therefore, $f$ evolves on each energy hypersurface.   
\item  two-body relaxation.  It operates on a time scale of
order $\displaystyle{\frac{N}{\ln N}t_{cr}}$, where $N$ is the number
of particles.  Though this {\sl two-body relaxation timescale}, $t_r$, is much longer than any other timescale we
have considered, this process leads to evolution of $f(E)$, and it
dominates the long-term evolution of large $N$-body systems.  It is
usually modelled by adding a {\sl collision term} of Fokker-Planck
type on the right side of eq.(\ref{eq:vlasov}).  

In this case the only
equilibrium solutions in a steady potential are those in which $f(E) \propto \exp(-\beta
E)$, where $\beta$ is a constant.  Then eq.(\ref{eq:Poisson}) becomes
Liouville's equation, and for the case of spherical symmetry the
relevant solutions are those corresponding to the {\sl isothermal sphere}.
\end{enumerate}

\subsection{Collisional equilibrium}

\begin{figure}
\vskip4truein
\includegraphics{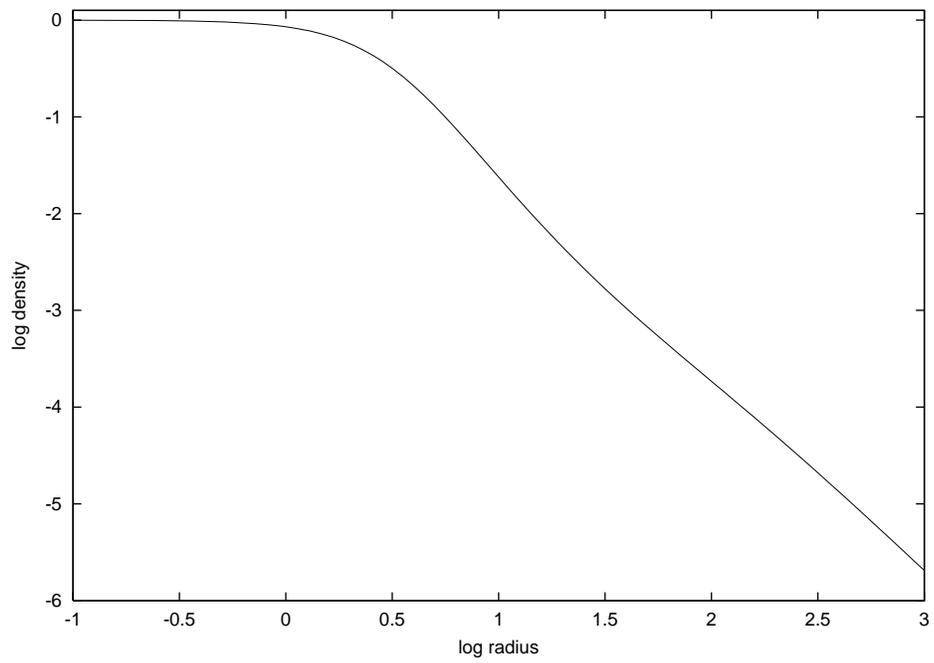}
\caption{The density profile of the non-singular isothermal model,
with conventional scalings.}
\label{fig:isothermal}
\end{figure}

We consider the collisional evolution of an $N$-body system further in
Sec.\ref{sec:collisional_evolution}, and here develop the fundamental
ideas about the isothermal model.
The isothermal model has infinite mass, and much has been learned by
considering a model confined within an adiabatic
boundary or enclosure.  There is a series of such models,
characterised by a single dimensionless parameter, which can be
taken to be the ratio between the central density and the density at
the boundary, $\rho_0/\rho_e$  (Fig.\ref{fig:isothermal}).

These models are extrema of the Boltzmann entropy $S = - k \int f\ln
fd^6\tau$, where $k$ is Boltzmann's constant, and the integration is
taken over all available phase-space.  Their stability may be
determined by evaluating the second variation of $S$.  It is found
that it is negative definite, so that $S$ is a local maximum and the
configuration is stable, only if $\rho_0/\rho_e < 709$ approximately.
A physical explanation for this is the following.  In the limit when
$\rho_0/\rho_e\simeq1$ the self-gravity (which causes the spatial
inhomogeneity) is weak, and the system behaves like an ordinary
perfect gas.  When $\rho_0/\rho_e\gg1$, however, the system is highly
inhomogeneous, consisting of a core of low mass and high density
surrounded by an extensive halo of high mass and low density.
Consider a transfer of energy from the deep interior to the envelope.
In the envelope, which is restrained by the enclosure, the additional
energy causes a rise in temperature, but this is {\sl small}, because of the
huge mass of the halo.  Extraction of energy from around the core,
however, causes the bodies there to sink and accelerate; and, because
of the negative specific heat of a self-gravitating system, they gain
more kinetic energy than they lost in the original transfer.  Now the
system is hotter in the core than in the halo, and the transfer of
energy from the interior to the exterior is self-sustaining, in a {\sl
gravothermal runaway}.  The isothermal model with large density
contrast is therefore unstable.

The negative specific heat, and the lack of an equilibrium which
maximises the entropy, are two examples of the anomalous thermodynamic
behaviour of  the self-gravitating $N$-body problem.  They are related
to the long-range nature of the gravitational interaction, the
importance of boundary terms, and the non-extensivity of the energy.
Another consequence is the inequivalence of canonical and
microcanonical ensembles.  

\subsection{Numerical methods}

The foregoing considerations are difficult to extend to systems
without a boundary, though they are a vital guide to the behaviour
even in this case.  Our knowledge of such systems is due largely to
numerical experiments, which fall into several classes:
\begin{enumerate}
\item Direct $N$-body calculations.  These minimise the number of
simplifying assumptions, but are expensive.  Special-purpose hardware
is readily available which greatly accelerates the necessary
calculations.  Great care has to be taken in the treatment of few-body
configurations, which otherwise consume almost all resources.
\item Hierarchical methods, including tree methods, which shorten the
calculation of forces by grouping distant masses.  They are mostly
used for collisionless problems.
\item Grid-based methods, which are used for collisionless problems
\item Fokker-Planck methods, which usually require a theoretical knowledge of
the statistical effects of two-, three- and four-body interactions.
Otherwise they can be very flexible, especially in the form of Monte
Carlo codes.
\item Gas codes.   The behaviour of a self-gravitating
system is simulated surprisingly well by modelling it as a
self-gravitating perfect gas, rather like a star.  
\end{enumerate}

\subsection{Collisional evolution}\label{sec:collisional_evolution}

Consider an isolated $N$-body system, which we suppose initially to be
given by a spherically symmetric equilibrium solution of
eqs.(\ref{eq:vlasov}-\ref{eq:Poisson}), such as
eq.(\ref{eq:plummer}).  The temperature decreases with increasing
radius, and a gravothermal runaway causes the ``collapse'' of the
core, which reaches extremely high density in finite time.  (This
collapse takes place on the long two-body relaxation time scale, and
so it is not the rapid collapse, on a free-fall time scale, which the
name rather suggests.)  

At sufficiently high densities the time scale of three-body reactions
becomes competitive.  These create bound pairs, the excess energy
being removed by a third body.  From the point of view of the
one-particle distribution function, $f$, these reactions are
exothermic, causing an expansion and cooling of the high-density
central regions.  This temperature inversion drives the gravothermal
runaway in reverse, and the core expands, until contact with the cool
envelope of the system restores a normal temperature profile.  Core
collapse resumes once more, and leads to a chaotic sequence of
expansions and contractions, called {\sl gravothermal oscillations}.

The monotonic addition of energy during
the collapsed phases causes a secular expansion of the system, and a
general increase in all time scales.  In each relaxation time a small
fraction of the masses escape, and eventually (it is thought) the
system consists of a dispersing collection of mutually unbound single
masses, binaries and (presumably) stable higher-order systems.  

It is very remarkable that the long-term fate of the largest
self-gravitating  $N$-body systems appears to be intimately linked with
the three-body problem.

\section{Further Reading}

\parindent=0pt


Arnold, V.I., Kozlov, V.V., Neishtadt, A.I., 1993, {\sl Mathematical
  Aspects of Classical and Celestial Mechanics}, 2nd ed, (Berlin: Springer)

Barrow-Green, J., 1997, {\sl Poincar\'e and the three body problem}, 
(Providence: American Mathematical Society)

Binney, J., Tremaine, S., 1988, {\sl Galactic Dynamics} (Princeton:
Princeton University Press)

Chenciner A., Montgomery R., 2000, A Remarkable Periodic Solution of
the Three-Body Problem in the Case of Equal Masses, {\sl Ann. Math.},
{\bf 152}, 881-901

Dauxois, T., et al (eds), 2003, {\sl Dynamics and Thermodynamics of
Systems with Long Range Interactions} (Berlin: Springer-Verlag)

Diacu, F., 2000, A Century-Long Loop, {\sl Math. Intelligencer}, {\bf}
22, 2, 19-25

Heggie, D.C., Hut, P., 2003, {\sl The Gravitational Million-Body
Problem} (Cambridge: Cambridge University Press)

Marchal, C., 1990, {\sl The Three-body Problem}
(Amsterdam: Elsevier)

Murray, C.D., Dermott, S.F., 2000, {\sl Solar System Dynamics},
(Cambridge: Cambridge University Press)

Padmanabhan, T., 1990, Statistical mechanics of gravitating systems,
{\sl Phys. Rept.} {\bf 188}, 285-362

Siegel, C.L., Moser, J.K.,  Kalme, C.I., 1995, {\sl  Lectures on
Celestial Mechanics} (Berlin:
Springer-Verlag)

Steves, B.J.,
Maciejewski, A.J. (eds), 2001, {\sl The Restless Universe.
  Applications of Gravitational $N$-Body Dynamics to Planetary,
  Stellar and Galactic Systems} (Bristol:  IoP and SUSSP)

Szebehely, V.,  1967, {\sl Theory of orbits : the restricted problem
of three bodies} (New York : Academic Press)

\section{Figure Captions}
\newpage
\newpage
\newpage

\begin{figure}
\vskip4truein
\includegraphics{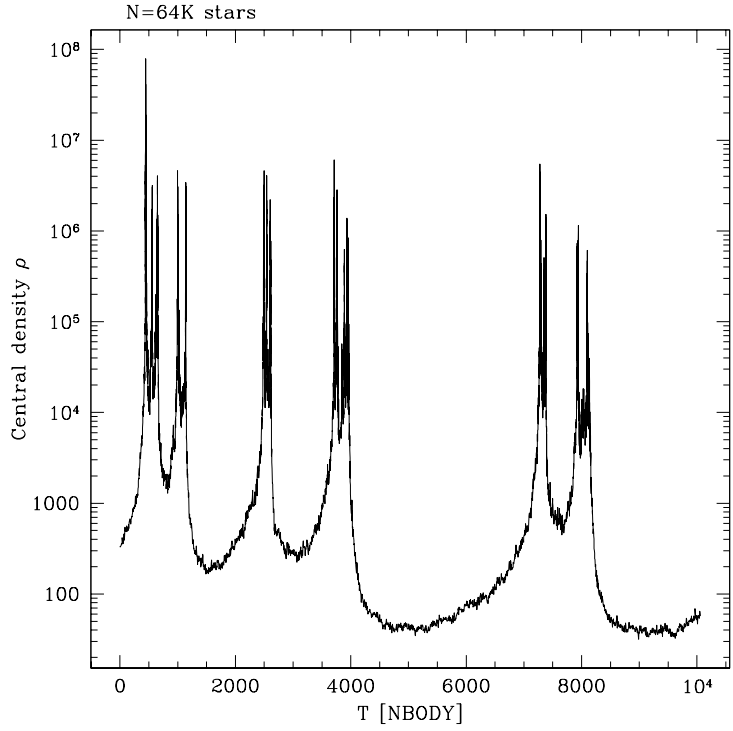}
\caption{Gravothermal oscillations in an  $N$-body system with $N = 65536$.  The
central density is plotted as a function of time, in units such that
$t_{cr}=2\sqrt{2}$.  
{\sl Source: H. Baumgardt, P. Hut, J. Makino, with permission}}
\label{fig:gto}
\end{figure}

\vfill

\end{document}